\documentclass[12pt,preprint]{aastex}

\slugcomment{This manuscript is accepted by \textbf{Ap\&SS}}

\shorttitle{Accretion disk and landing LMCs}

\shortauthors{M.elyasi & M.Nejad-Asghar}

\begin{document}

\title{Occurrence of instability through the protostellar accretion disks by landing of low-mass condensations}

\author{Mahjubeh Elyasi, Mohsen Nejad-Asghar}

\affil{Department of Atomic and Molecular Physics, University of
Mazandaran, Babolsar, Iran}

\email{nejadasghar@umz.ac.ir}

\begin{abstract}
Low-mass condensations (LMCs) are observed inside the envelope of
the collapsing molecular cloud cores. In this research, we
investigate the effects of landing LMCs for occurrence of
instability through the protostellar accretion disks. We consider
some regions of the disk where duration of infalling and landing of
the LMCs are shorter than the orbital period. In this way, we can
consider the landing LMCs as density bumps and grooves in the
azimuthal direction of an initial thin axisymmetric steady state
self-gravitating protostellar accretion disk (nearly Keplerian).
Using the linear effects of the bump quantities, we obtain a
characteristic equation for growth/decay rate of bumps; we
numerically solve it to find occurrence of instability. We also
evaluate the minimum-growth-time-scale (MGTS) and the enhanced mass
accretion rate. The results show that infalling and landing of the
LMCs in the inner regions of the protostellar accretion disks can
cause faster unstable modes and less enhanced accretion rates
relative to the outer regions. Also, more fragmentation of landed
LMCs in the azimuthal direction have less chance for instability,
and then can produce more values of enhanced mass accretion rate.
\end{abstract}


\section{Introduction}
Dense cores through the molecular clouds are nurseries of
protostellar accretion disks. There has been a lot of observations
which show that the structure of these dense cores are clumpy with
small masses and sizes. For example, Langer et al.~(1995) observed
low mass condensations (LMCs) in the core D of Taurus Molecular
Cloud~1 in the regime of 0.007-0.021 pc and $0.01-0.15M_{\bigodot}$.
Also, we can refer to the discovery of very low luminosity objects
by the Spitzer From Molecular Cores to Planet-Forming disks' (c2d)
project (Lee et al.~2009, Dunham et al.~2014) or the results gained
by Launhardt et al.~(2010), in which they found that at least
two-thirds of 32 studied isolated star-forming cores show evidence
of forming multiple stars. For other observations, we can refer to
millimeter and submillimeter observations toward two prestellar
cores SM1 and B2-N5, in active cluster forming regions, that showed
the presence of small scale objects inside the prestellar cores with
masses in the range of $10^{-2}-10^{-1}$ $M_{\odot}$ and sizes of a
few hundred AU (Nakamura et al. 2012). These observations, and also
the works of Pirogov \& Zinchenko~(2008) and Tachihara~(2013), can
be used as a witness to the existence of LMCs in the molecular cloud
cores. Since the masses of these condensations are small and the
molecular cloud cores are almost quiescent, gravitational
instability and turbulent can not be considered as the responsible
mechanisms for clumping through the cores. Nejad-Asghar(2011a)
showed that thermal instability may be considered as an important
mechanism for formation of the LMCs in the envelope of the molecular
cloud cores.

Initial rotation of the collapsing dense cores can lead to the
formation of an accretion disk, in which may ultimately be ended to
the formation of the proto-planetary entities (e.g., Stahler~2004,
Hartmann~2009). There are two known mechanisms for the formation of
the condense proto-planetary entities through the protostellar
accretion disks: the core accretion (Goldreich \& Ward~1973, Miguel
\& Brunini~2009), and the disk instability (Toomre~1964, Boss~1997,
Zhu et.al.~2012). The core accretion mechanism occurs from the
collision and coagulation of dusty solid particles into gradually
larger bodies until a massive enough proto-planet is formed. In the
instability model, a disk around a protostar may be fragmented by
instability and these pieces may become proto-planetary entities. In
this paper, we neglect the dusty solid particles and focus on the
latter mechanism. In order to occur the instability and disk
fragmentation, some physical conditions must be satisfied. Two
criteria are mostly used to discuss whether a protostellar disk is
likely to fragment. The first is Toomre's stability criterion
(Toomre 1964): $Q=\frac{c_{s}\kappa}{\pi G \Sigma}>1$, where $c_{s}$
is the sound speed, $\kappa$ is the local epicyclic frequency, and
$\Sigma$ is the disk surface density. If $Q<1$, the disk may be
gravitationally unstable (e.g., Binney and Tremaine~2008). The
second criterion is Gammie's cooling criterion (Gammie 2001):
$t_{cool} <\frac{1}{\Omega}$, where $\Omega$ is the angular
velocity. Gammie suggested that the disk must be quickly cooled
until the disk can fragment and gravitationally bound gaseous
objects form.

In this research, we consider the phase of the disk, in which the
matters from envelope (especially LMCs) of the collapsing molecular
cloud core are still infalling onto the disk. The high velocity of
infalling LMCs from the envelope of the cores onto the accretion
disk, leads to formation of the shocked waves (Mendoza 2009,
Nejad-Asghar 2011b). If the timescales of cooling and relaxing of
the shocked waves are much shorter than the orbital period, we
expect to form density bumps and grooves through the protostellar
accretion disks. Nejad-Asghar(2011c) showed that timescales for
formation shock waves and cooling down to form density bumps on the
accretion disks, are about few hundred years. In this way, we
consider some regions of the protostellar accretion disk where
infalling and landing of the LMCs on there can lead to formation of
density bumps and grooves in a timescales much shorter than orbital
period. These density bumps change the surface density of the
protostellar accretion disks and can affect on the gravitational
potential and pressure gradient, so that the disk may be unstable.
In this research, we consider a thin axisymmetric steady state
self-gravitating accretion disk in cylindrical coordinate and
investigate occurrence of instability caused by these density bumps
and grooves. Formulation of the problem is given in the section~2,
and the results with some concluding remarks are presented in the
section~3.

\section {Formulation of the problem}
The fundamental equations governing thin rotating disk with
$\textbf{u}=u_{R}\hat{r}+u_{\varphi}\hat{\varphi}$, in the
cylindrical polar coordinates $(R,\varphi)$ are (e.g., Clark $\&$
Carswell~2007)
\begin{equation}\label{mass}
    \frac{\partial \Sigma}{\partial t}+\frac{1}{R}\frac{\partial}{\partial R}(R \Sigma u_{R})+
    \frac{1}{R}\frac{\partial}{\partial \varphi}(\Sigma u_{\varphi})=0,
\end{equation}
\begin{equation}\label{moment1}
    \frac{\partial u_{R}}{\partial t}+u_{R}\frac{\partial u_{R}}{\partial
    R}+\frac{u_{\varphi}}{R}\frac{\partial u_{R}}{\partial \varphi}-\frac{u^{2}_{\varphi}}{R}=-\frac{\partial\Phi}{\partial R}
    -\frac{1}{\Sigma}\frac{\partial P}{\partial R},
\end{equation}
\begin{equation}\label{moment2}
    \frac{\partial u_{\varphi}}{\partial t}+u_{R}\frac{\partial u_{\varphi}}{\partial
    R}+\frac{u_{\varphi}}{R}\frac{\partial u_{\varphi}}{\partial \varphi}+\frac{u_{\varphi}u_{R}}{R}=
    -\frac{1}{R}\frac{\partial \Phi}{\partial \varphi}-\frac{1}{R\Sigma}\frac{\partial P}{\partial \varphi},
\end{equation}
where $\Sigma$ and $\Phi$ are the surface density and the
gravitational potential, respectively, and $P$ is the gas pressure
that is given by the equation of state
\begin{equation}\label{state}
    P=\frac{k_{B}}{\mu m_{H}}\Sigma T,
\end{equation}
where $k_{B}$, $\mu$, $m_{H}$, and $T$ are the Boltzman constant,
mean molecular mass, hydrogen atomic mass, and the temperature of
the disk, respectively. In this paper, we aim to study the effect of
the landing LMCs on the instability of the disk with initial mass
accretion rate less than $10^{-6}M_{\odot}yr^{-1}$. Since infalling
and landing of the LMCs on the accretion disk, produces shock waves
and viscous heating, we consider the thermal effect of viscosity,
while its dynamical effects in the momentum equations
(\ref{moment1}) and (\ref{moment2}) are ignored. It should be noted
that to analyze the dynamical effects in low-rate accretion flows
($\lesssim 10^{-6}M_{\odot}yr^{-1}$), we assume the viscous terms
are negligible compared to the potential and pressure gradient.
Also, we suppose the disk to be in thermal equilibrium, so that
energy sources and sinks are in local balance; the heating sources
include viscous dissipation and external irradiation intercepted by
the disk, and the energy sink is affected by radiative cooling.
Here, we do not directly evaluate the temperature and surface
density of the viscous heated accretion disks. Instead, we use a
low-rate accretion disk model with $\Sigma_0(R)$ and $T_0(R)$ as
recently obtained by the work of Rafikov~(2015).

We consider a system including a central protostar with mass
$M_{\ast}=M_{\odot}$ which is surrounded by a self-gravitating
gaseous disk (initial smooth accretion disk). Also, we assume the
disk is initially in steady state with axisymmetric behavior
$(\frac{\partial}{\partial t}=0~$\&$
~\frac{\partial}{\partial\varphi}=0)$. Also, in the low-rate
accretion flows, the radial velocity is smaller than the azimuthal
velocity, thus, the radial velocity is supposed to be negligible,
i.e., $u_{R}=0$. In this steady axisymmetric state, the equations
(\ref{mass})-(\ref{moment2}) reduce to
\begin{equation}\label{moment11}
    \frac{u^{2}_{\varphi}}{R}=-g+\frac{1}{\Sigma_0}\frac{dP_0}{dR},
\end{equation}
where, $g=-\frac{\partial\Phi}{\partial R}$ is the gravitational
field. We separate the gravitational field to the central mass
contribution, $g^{c}=-\frac{GM_{\ast}}{R^{2}}$, and the disk
contribution, which depends on the surface density through the
Poisson's equation (e.g., Hure $\&$ Pierens ~2005) given by
\begin{equation}\label{gdisk}
  g^{d}(R)=-G\int_{R_{in}}^{R_{out}}
  \sqrt{\frac{R'}{R}}\frac{\kappa}{R}\Sigma_0(R')[K(\kappa)-\frac{E(\kappa)}{\overline{\omega}}]dR'.
\end{equation}
where, $K(\kappa)$ and $E(\kappa)$ are the complete elliptic
integrals of the first and second kinds, respectively,
$\kappa=2\frac{\sqrt{R'R}}{(R'+R)}$ is their modulus, and
$\overline{\omega}=\frac{(R'-R)}{(R'+R)}$. Substituting
$u_{\varphi}=R\Omega(R)$ into equation (\ref{moment11}) and using
equation (\ref{state}), the angular velocity is given by
\begin{equation}\label{omega2}
    \Omega_0=\sqrt{\Omega^{2}_{K}+\frac{1}{R}(-g^{d}+\frac{k_{B}}{\mu
    m_{H}}\frac{1}{\Sigma_0}\frac{d}{dR}(\Sigma_0
    T_0)}),
\end{equation}
where, $\Omega_{K}=\sqrt{\frac{GM_{\ast}}{R^{3}}}$ is the Keplerian
angular velocity.

\begin{figure}
\epsscale{.50} \center \plotone{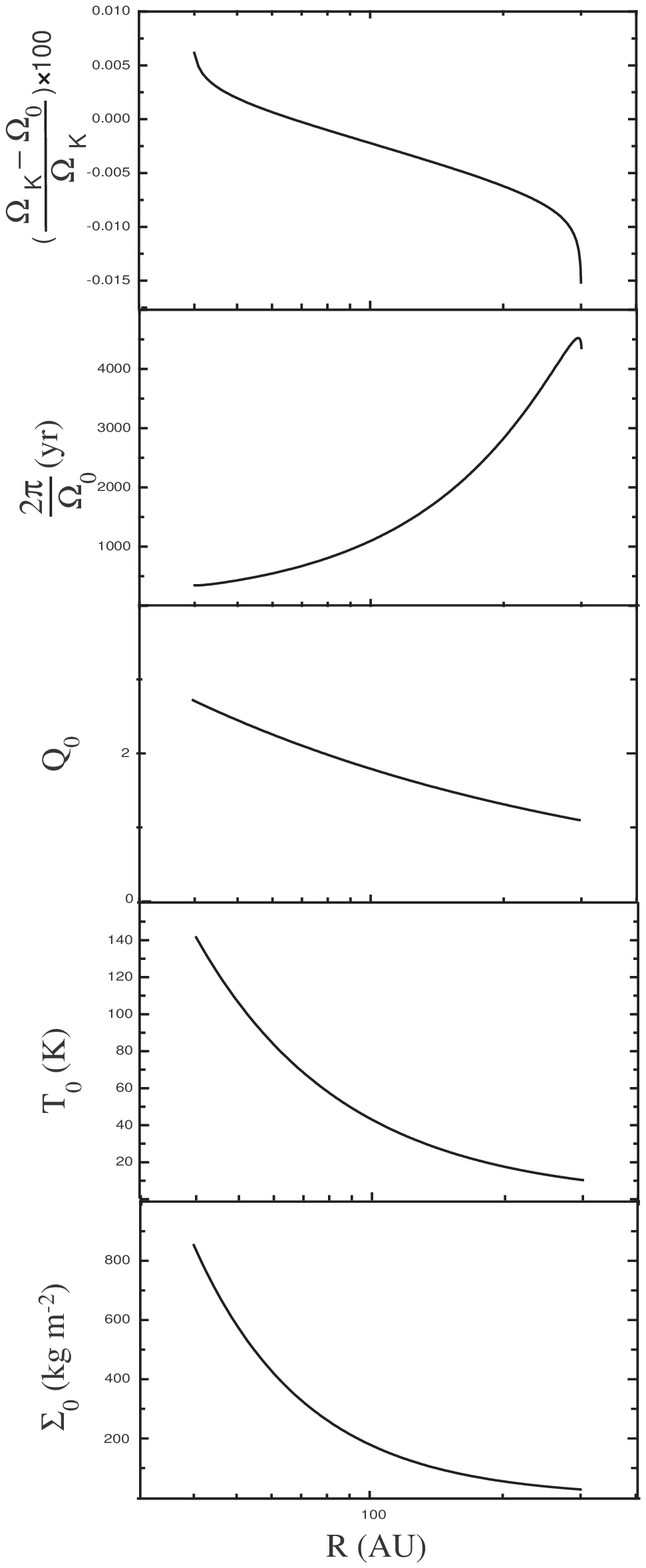} \caption{Radial profiles
of surface density $\Sigma_0$, temperature $T_0$, Toomre parameter
$Q_0=c_{s}\Omega_0/\pi G \Sigma_0$, orbital period $2\pi/\Omega_0$,
and relative difference between the angular velocity with its
Keplerian one $(\Omega_K-\Omega_0)/\Omega_K$, for initial
protostellar accretion disk model described by equation
(\ref{sigma0}). \label{omega}}
\end{figure}

We choose the dimensionless scales for length, mass, time, and
temperature as $[R]=100 AU$, $[M]=2.9\times 10^{25} kg$, $[t]=50yr$,
and $[T]=10 K$, respectively, so that, $[{\Sigma}]=0.7 [M][R]^{-2} =
0.09 kg/m^{2}$, $G=1.23\times10^{-6} [R]^3[M]^{-1}[t]^{-2}$, and
$\frac{k_{B}}{\mu m_{H}}=5.73\times10^{-4}[R]^{2}[T]^{-1}[t]^{-2}$
with choosing $\mu=1.5$. With these scales, we consider a model for
initial smooth low-rate accretion disk ($\approx
7\times10^{-7}M_{\odot}yr^{-1}$) with surface density and
temperature profiles as
 \begin{equation}\label{sigma0}
    {\Sigma}_0(R)=1995.2 R^{-1.7},~~~T_0(R)=4.3R^{-1.3},
 \end{equation}
respectively (Rafikov~2015). Fig.~\ref{omega} shows the radial
profiles of the surface density $\Sigma_0$, temperature $T_0$,
Toomre parameter $Q_0=c_{s}\Omega_0/\pi G \Sigma_0$, orbital period
$2\pi/\Omega_0$, and relative difference between the angular
velocity with its Keplerian one $(\Omega_K-\Omega_0)/\Omega_K$. We
can see that the disk rotation is nearly Keplerian. According to the
work of Nejad-Asghar~(2011c) for the timescales for formation of
density bumps (i.e., few hundred years), we consider regions of the
protostellar disk (i.e., radius between $0.4$ to $3[R]$), which the
orbital period is greater than these timescales.

Here, we assume that density bumps are small and can cause to
produce the small quantities: gravitational potential ($\Phi_{1}$),
pressure ($P_{1}$), radial velocity ($u_{R1}$), and azimuthal
velocity ($u_{\phi1}$). For finding the relation between these
quantities to the density bump, we consider
${\Sigma}={\Sigma}_{0}+{\Sigma}_{1}$, $u_{R}=u_{R1}$,
$u_{\varphi}=u_{\varphi0}+u_{\varphi 1}$, $P=P_{0}+P_{1}$, and
${\Phi}={\Phi}_{0}+{\Phi}_{1}$, where the subscripts "0" and "1"
represent the initial steady state accretion disk and bump
quantities, respectively. Substituting these relations into
equations (\ref{mass})-(\ref{moment2}), in the first-order
approximation, we have
\begin{equation}\label{perturbmass}
    \frac{\partial \Sigma_{1}}{\partial t}+\frac{1}{R}\frac{\partial}{\partial
    R}(R\Sigma_{0}u_{R1})+\frac{\Sigma_{0}}{R}\frac{\partial}{\partial\varphi}(u_{\varphi1})+\Omega_{0}\frac{\partial}{\partial\varphi}(\Sigma_{1})=0,
\end{equation}
\begin{equation}\label{perturbmoment1}
    \Sigma_{0}\frac{\partial u_{R1}}{\partial t}+\Sigma_{0}\Omega_{0}\frac{\partial
    u_{R1}}{\partial\varphi}-2\Sigma_{0}\Omega_{0}u_{\varphi1}-\Sigma_{1}R\Omega^{2}_{0}=-\Sigma_{0}\frac{\partial\Phi_{1}}{\partial R}
    -\Sigma_{1}\frac{\partial\Phi_{0}}{\partial R}-\frac{\partial P_{1}}{\partial R},
\end{equation}
\begin{equation}\label{perturbmoment2}
    \Sigma_{0}\frac{\partial u_{\varphi1}}{\partial
    t}+\Sigma_{0}u_{R1}\frac{\partial u_{\varphi0}}{\partial R}+\frac{1}{R}\Sigma_{0}u_{\varphi0}\frac{\partial u_{\varphi1}}{\partial\varphi}
    +\Sigma_{0}\frac{u_{\varphi0}u_{\varphi1}}{R}=-\frac{\Sigma_{0}}{R}\frac{\partial\Phi_{1}}{\partial\varphi}-\frac{1}{R}\frac{\partial
    P_{1}}{\partial\varphi}.
\end{equation}
The azimuthal and time dependency of the bump quantities can be
expanded by the Fourier terms as follows
\begin{eqnarray}\label{bumpequ}
&&~~~~ \nonumber  u_{R1}=Re[u_{Ra}(R)\exp(i(m\varphi-\omega t))],~~~~  u_{\varphi 1}=Re[u_{\varphi a}(R)\exp(i(m\varphi-\omega t))],\\
&&~~~~  \Sigma_{1}~=Re[\Sigma_{a}(R)\exp(i(m\varphi-\omega t))] ,~~~~~~ \Phi_{1}=Re[\Phi_{a}(R)\exp(i(m\varphi-\omega t))],\\
&&~~~~\nonumber P_{1}=Re[P_{a}(R)\exp(i(m\varphi-\omega t))],
\end{eqnarray}
where, $m$ is the azimuthal mode number and $\omega$ is a complex
number which its imaginary part indicates growth/decay rate
according to its sign. In general, $\omega$ must be a function of
bump position in the accretion disk, but for simplicity, we consider
it as a constant value at the peak-point of the bump. Here, we
assume that infalling and landing of the LMcs lead to formation
Gaussian density bumps at the timescale shorter than orbital period,
as
\begin{equation}\label{bumpdensity}
    {\Sigma}_{a}(R)={\Sigma}'_{a}exp(-4(\frac{R-{R}_{1}}{\Delta_{1}})^{2}),
\end{equation}
where $\Sigma'_{a}$, $R_{1}$, and $\Delta_{1}$ are the amplitude,
peak-point position, and e-folding width of the bump, respectively.
Also, we assume that the bumped gravitational potential
$\Phi_{a}(R)$ is related to the density bump via the Poisson
equation (Hure $\&$ Pierens 2005)
\begin{equation}\label{gdiska}
  \Phi_{a}(R)=-2G\int_{R_{in}}^{R_{out}}
  \sqrt{\frac{R'}{R}}{\kappa}\Sigma_{a}(R')K(\kappa)dR',
\end{equation}
and the bumped pressure is given by $ P_{a}=\frac{k_{B}}{\mu
m_{H}}\Sigma_{a}T_{0}$.

Now, considering equations (\ref{bumpequ}), the equations
(\ref{perturbmass})-(\ref{perturbmoment2}) reduce to
\begin{equation}\label{13}
    i(m\Omega_{0}-\omega)\Sigma_{a}+\frac{1}{R}\frac{d}{dR}(R\Sigma_{0}u_{Ra})+\frac{im\Sigma_{0}}{R}u_{\varphi
    a}=0,
\end{equation}
\begin{equation}\label{14}
   i(m\Omega_{0}-\omega)\Sigma_{0}u_{Ra}+(\frac{d\Phi_{0}}{dR}-R\Omega_{0}^{2})\Sigma_{a}-2\Sigma_{0}\Omega_{0}u_{\varphi_{a}}=
   -\Sigma_{0}\frac{d\Phi_{a}}{dR}-\frac{dP_{a}}{dR},
\end{equation}
\begin{equation}\label{15}
  i(m\Omega_{0}-\omega)\Sigma_{0}u_{\varphi_{a}}+\Sigma_{0}\frac{d}{dR}(R\Omega_{0}+\Sigma_{0}\Omega_{0})u_{Ra}=
  -i\frac{m\Sigma_{0}}{R}\Phi_{a}-i\frac{m}{R}P_{a}.
\end{equation}
By merging equations (\ref{14}) and (\ref{15}), the amplitude of
radial and azimuthal components of the bumped velocity are obtained
as follows
\begin{equation}\label{amp_radial}
    u_{Ra}=-i\frac{((R\Omega_{0}^{2}-\frac{d\Phi_{0}}{dR}-\frac{1}{\mu}\frac{dT_{0}}{dR}-2\frac{m\Omega_{0}T_{0}}{R\mu} \frac{1}{(m\Omega_{0}-\omega)})\Sigma_{a}
    -\Sigma_{0}\frac{d\Phi_{a}}{dR}-\frac{T_{0}}{\mu}\frac{d\Sigma_{a}}{dR}-2\frac{m\Omega_{0}\Sigma_{0}}{R}\frac{\Phi_{a}}{(m\Omega_{0}-\omega)} )}
    {\Sigma_{0}((m\Omega_{0}-\omega)-2\Sigma_{0}\Omega_{0}
    \frac{(R\frac{d\Omega_{0}}{dR}+2\Omega_{0})}{(m\Omega_{0}-\omega)})},
\end{equation}
\begin{equation}\label{amp_azimuthal}
 u_{\varphi a}=i\frac{1}{(m\Omega_{0}-\omega)\Sigma_{0}}(\Sigma_{0}\frac{d}{dR}(R\Omega_{0}
 +\Sigma_{0}\Omega_{0})u_{Ra}+i\frac{m\Sigma_{0}}{R}\Phi_{a}+i\frac{m}{R}P_{a}).
\end{equation}
Now, by substitution $u_{Ra}$ and $u_{\varphi a}$ into equation
(\ref{13}), we obtine a characteristic polynomial equation as
\begin{equation}\label{diffusion}
  a_{6}\omega^{6}+a_{5}\omega^{5}+a_{4}\omega^{4}+a_{3}\omega^{3}+a_{2}\omega^{2}+a_{1}\omega+a_{0}=0,
\end{equation}
where the coefficients $a_{0}$, $a_{1}$, $a_{2}$, $a_{3}$, $a_{4}$,
$a_{5}$, and $a_{6}$ depend on the radial profile of surface
density, temperature and angular velocity of our chosen model for
smooth protostellar accretion disk (i.e., $\Sigma_{0}$, $T_{0}$ and
$\Omega_{0}$), on the density bump parameters (i.e., $\Sigma'_{a}$
and $\Delta_{1}$ and its peak-point position $R_{1}$), and
especially on azimuthal mode number $m$, which represents initial
fragmentation of bump quantities (number of bumps and grooves)
across the azimuthal direction at radius $R_1$ of the disk.

\section{Numerical results and conclusion}
In this paper, a phase of collapsing molecular cloud cores was
surveyed, in which an accretion disk around a protostar was formed.
In this phase, the infalling and landing of LMCs, which were
preformed in the core envelope, could lead to change the dynamics of
the protostellar disk. Infalling and landing of LMCs on the
protostellar accretion disk produces shocked waves, which if they
cool rapidly in a timescale shorter than the orbital period, can
lead to formation of density bumps and grooves through initial
smooth disk. Assuming the bump quantities are small in amplitude,
and using the linear analysis on the basic equations, we obtained
the characteristic equation (\ref{diffusion}).

The characteristic equation (\ref{diffusion}) is a polynomial
function of $\omega$, and its coefficients are evaluated at the peak
point position of the density bumps, $R_1$. We use the Laguerre's
method (Press et.al~1992) to find the roots of this polynomial
equation. The positive/negative sign of the imaginary part of
$\omega$ represents the growth/damping rate. We locate the density
bumps at different radii of the disk, and find the roots of the
characteristic equation. According to the obtained imaginary part of
the root, the instability and growth rates of the bumped accretion
disk at different radii can be found. If at each radius, we obtain
many roots with positive imaginary parts, we consider maximum value
of them as maximum growth rate. Inverse of the maximum growth rate,
$\frac{1}{\max(\Im(\omega))}$, is the minimum-growth-time-scale
(MGTS), which represent the growth time of the bump quantities and
formation of e-folded density clumping through protostellar
accretion disks.

\begin{figure}
\epsscale{.70} \center \plotone{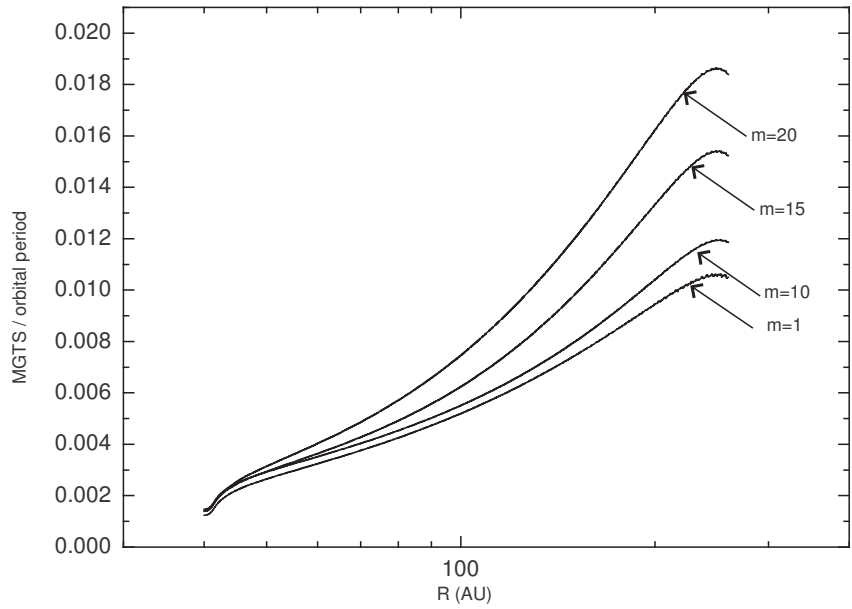} \caption{The
normalize MGTS by the orbital period, as a function of the position
of the pick-point of bumps, for azimuthal mode numbers $m=1,10,15,$
and $20$. The parameters of density bump are chosen as
$\Sigma'_{a}=0.01\Sigma_{0}$ and $\Delta_{1}=0.1 [R]= 10
AU$.\label{Time}}
\end{figure}

Considering density bumps with parameters
$\Sigma^{'}_{a}=0.01\Sigma_{0}$ and $\Delta_{1}=0.1[R]=10AU$, the
normalized MGTS by the orbital period, as a function of the position
of the pick-point of bump, $R_1$, are shown in the Fig.~\ref{Time},
for different values of azimuthal mode number $m=1,10,15$, and $20$.
As can be seen, increasing the azimuthal number $m$, causes the
normalized MGTS to increase. Increasing $m$ would produce more
values of bumps and grooves in the azimuthal direction. As a result,
if infalling and landing of the LMCs would lead to less
fragmentation of the bumps (i.e., smaller values of $m$), MGTSs
decrease and instability would happen with more chance. Moreover,
landing of LMCs in the outer parts of the accretion disk lead to
increasing of MGTSs. Therefore, bumping in the denser parts of the
protostellar disk (i.e., smaller values of $R$) would cause the
instability to happen faster. From the results, it can be said that
bumping closer to the protostar (i.e., smaller values of $R$) would
cause more instability, and thus increases the probability of
formation of proto-planetary entities over there.

\begin{figure}
\epsscale{.75} \center \plotone{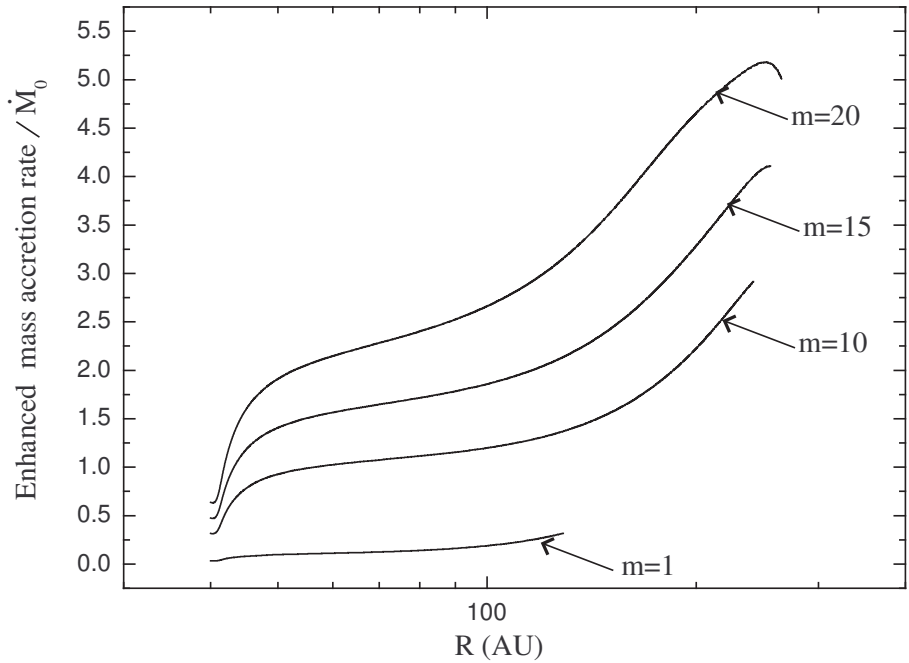} \caption{The ratio of
enhanced mass accretion rate to the initial mass accretion rate
$\dot{M}_0 \approx 7 \times 10^{-7} M_{\odot}yr^{-1}$, as a function
of the position of the pick-point of bumps, for azimuthal mode
number $m=1,10,15,$ and $20$. The parameters of density bump are
chosen as $\Sigma'_{a}=0.01\Sigma_{0}$ and $\Delta_{1}=0.1 [R] =
10AU$. \label{Mdot}}

\end{figure}

The enhanced mass accretion rates, $\dot{M}=2\pi R\Sigma_{0}
|u_{Ra}|$, relative to the initial mass accretion rate $\dot{M}_0
\approx 7 \times 10^{-7} M_{\odot}yr^{-1}$, are shown in the
Fig.~\ref{Mdot}, as a function of the position of the pick-point of
bumps, for $m=1,10,15$, and $20$. It is seen that, mass accretion
rate is smaller in the inner regions of the protostellar disk where
the normalized MGTSs have smaller values. In regions where
instability would be more/less likely to occur (i.e.,
smaller/greater values of MGTS), the rate of mass accretion onto the
central mass object will be reduced/incremented via
increasing/decreasing of coagulation of matter and growth of
condensations. In the other words, when the MGTS is short, formed
bump can grow and get bigger. Thus, the matters will be gathered in
the place of the formed density bumps and cannot fall down on the
central mass, therefore enhanced mass accretion rate decreases, and
vise versa. Also, the results of Fig.~\ref{Mdot} show that the
enhanced mass accretion rate increases with increasing the azimuthal
mode number $m$. Thus, more fragmentation of landed LMCs in the
azimuthal direction (i.e., greater values of $m$) have less chance
for instability, and then can produce more values of enhanced mass
accretion rate. In other words, smaller values of azimuthal mode
number (i.e., smaller numbers of bumps and grooves across the
azimuthal direction), in the inner regions of the protostellar
accretion disks, have more chance for instability and growing of
condensations to form proto-planetary entities.




\clearpage

\clearpage

\end{document}